\documentclass[preprint,times]{aastex631}

\usepackage{amsmath}
\usepackage{multirow}
\usepackage{graphicx}
\received{November, 12, 2025}
\revised{July, 9, 2026}
\accepted{July, 13, 2026}

\submitjournal{The Astrophysical Journal}

\shorttitle{Coronal $IQUD$ Magnetometry}
\shortauthors{Paraschiv}
\graphicspath{{./}{figures/}}
\begin{document}
\fontsize{10}{12}\selectfont
\title{Inferring 3D Coronal Magnetic Fields through Seismology-assisted Inversions of IQU-only Spectropolarimetric Observations}

%\correspondingauthor{August Muench}
%\email{greg.schwarz@aas.org, gus.muench@aas.org}

\author[0000-0002-3491-1983,gname=Alin,sname=Paraschiv]{Alin Razvan Paraschiv}
\affiliation{National Solar Observatory, 3665 Discovery Drive, Boulder CO 80303, USA}
%\email{arparaschiv ``at'' nso.edu}

\begin{abstract}
%% Topic
The routine measurement of the Stokes $IQUV$ signals of emission lines in the solar corona is still a challenging endeavor, particularly for observations using small-aperture instruments.
%5 Methods and results
Therefore, recent studies have explored the use of coronal seismology and propagating Alfv\'enic waves as alternative diagnostics of the coronal magnetic field. In particular, Z. Yang et al. showed that plane-of-sky (POS) phase-speed measurements provide a direct and consistent diagnostic of the POS magnetic-field component (B$_{\text{POS}}$). Building on recent theoretical advances in solar coronal polarization, we devise a novel inversion scheme and extended the CLEDB software package to exploit Stokes $IQU$ observations of two coronal emission lines combined with coronal seismology-derived B$_{\text{POS}}$ estimations. Using this framework, we perform a comprehensive statistical analysis demonstrating that this combination of diagnostics allows us to infer both magnetic-field orientation and strength with an accuracy comparable to that of Stokes $IQUV$ spectropolarimetry. 
%% usefulness
This type of diagnostics is particularly suited for the \ion{Fe}{13} 1074.7 and 1079.8 nm line pair routinely observed by new-generation instruments such as DKIST Cryo-NIRSP, DL-NIRSP, MLSO CoMP/UCoMP, and targeted by the future COSMO and CORSAIR efforts.

\end{abstract}


\section{Introduction} \label{sec:intro}
%Purpose
Until recently, routinely measuring the full polarization state (Stokes $I,Q,U$ and $V$) of forbidden coronal emission lines was not reliably achieved by available instrumentation. As shown by the scarce, fairly-certain measurements presented in \citet{lin2000,lin2004}, measuring Stokes $V$ in the solar corona is not a trivial task. This, along with inherent assumptions, degeneracies, and sensitivity to observational and modeling assumptions led to concerns regarding the viability of interpreting and inverting coronal Stokes $IQUV$ polarization in general \citep{judge2013,Casini+2017,Dima+2020}. The recent results presented by \citet{schad2024} unequivocally demonstrate the coronal diagnostic potential and the viability of measuring Stokes $IQUV$ of near infrared (IR) lines, although more progress needs to be made in order for such measurements to become a routine observational capability. Previously, it has been shown that the information contained in polarized coronal observations allows for inferences of plasma properties, observational geometry, magnetic orientation and field strength \citep{Plowman2014,Dima+2020,par+judge2022}. In this work, we show that one can use the inversion methodology developed by \citet{par+judge2022} to infer magnetic orientation from Stokes $I$, $Q$, and $U$ components and couple that with the independent measure of magnetic field strength $B$ and orientation derived from coronal seismology methods \citep{tomczyk2007,morton+2016,yang2020+,yang2024} to infer vector coronal magnetic fields. This bypasses the need for including the Stokes $V$ measurements that often require much higher signal-to-noise (SNR) in observations. The inherent limitations, assumptions, and degeneracies will be discussed in the context of their impact on the reliability of the inferred solutions.

%%Coronal spectroscopy
The ground-based spectroscopic observation of important coronal IR lines such as \ion{Fe}{13} 1074.7 nm, \ion{Fe}{13} 1079.8 nm, \ion{Si}{10} 1430.1 nm, \ion{Mg}{8} 3028.3 nm, \ion{Si}{9} 3934.3 nm is subject to practical limitations, including scattered disk light and telluric absorption, as shown in the review by \citet{penn2014}. Theoretical calculation of coronal line-formation and expected intensities of these ions have been performed by \citet{Sahal1974,house1977,querfeld1982,Arnaud1987,judge1998}. \citet{Ali_2022} coupled ground-based spectral observations and theoretical line intensity calculations of coronal IR lines to discuss the most probable atmospheric absorption influences in select IR spectral regions where coronal ions form. High altitude detection of new and expected coronal lines have been presented by \citet{samra2018,samra2022,samra2025}, which further enhanced our understanding of highly ionized ions forming in the IR.  

%Coronal line theory and synthesis 
The theoretical framework of polarization in coronal IR lines has been gradually developed over the last decades. \citet{Casini+2017} present a detailed review of coronal ions and magnetic diagnostics. The above-mentioned lines are part of a class of forbidden transition magnetic dipole lines (M1), each with linear polarization (Stokes $Q$ and $U$) formed in the  strong-field (or \emph{saturated}) limit of the Hanle effect, and circular polarization (Stokes $V$) in the weak-field limit of the Zeeman effect \citep{Sahal1977,landi82,Casini+Judge1999}. The linear polarization gives access to the orientation of the magnetic field, while the circular polarization gives access to the magnetic field strength. Transverse coronal fields observed off-limb are prohibitively difficult to detect via the Zeeman effect, due to the quadratic dependence of the signal on field strength, as shown for example through the theoretical calculations of \citet{Supriya2025}. The theoretical formalism for coronal emission lines in saturated-Hanle and Zeeman regimes was developed by \citet{Casini+Judge1999} and \citet{lin2000b}, expanding earlier works \citep[e.g.][]{Sahal1977,landi82,landi91}. The formalism enabled the development of the Coronal Line Emission \citep[CLE;][]{Judge+casini2001} spectral synthesis code. The importance of accounting for Thompson scattering effects in coronal polarization has been subsequently discussed by \cite{Li2017}.  \citet{schad+dima2021} used a novel python based spectral synthesis and forward modeling tool \citep[PyCELp;][]{pycelp2021} to study anisotropy effects and atomic alignment in the presence of symmetry-breaking of the emergent polarized emission due to active regions.

%%Coronal magnetometry+wave propagation
\citet{judge2007} established that the IR coronal lines like \ion{Fe}{13} are rich in terms of information potential for coronal magnetic fields, but stresses that interpreting and inverting these will not be straightforward. Multiple works focusing on inferring coronal magnetic fields from polarized observations have been proposed, each having unique strengths and weaknesses. A series of inversion frameworks under the single-point approximation were proposed. This important assumption requires that the line-of-sight (LOS) integration of the observed optically-thin plasma is representative of a single, or at least a dominant homogeneous coronal structure, coherent in terms of its plasma, magnetic and geometrical parameters, and not a superposition of multiple disjointed coronal structures. Analytical inversions focusing on the Hanle and Zeeman effects have been developed by \citet{Plowman2014} and \citet{Dima+2020}. When including multiple vantage points or observations separated by long periods of time, the coronal field could be inverted using tomographic techniques \citep{Kramar2006,Kramar+2016}, but where more dynamic activity will be hard to recover. 

Significant degeneracies exist in both atomic properties and the level configuration of atomic calculations that contribute to interpreting the \ion{Fe}{13} line pair. \citet{schiffmann+2021} pursued detailed atomic Land\'{e} \emph{g} factor calculation for S-like and Si-Like atoms finding a deviation in the order of $\sim1\%$ from the standard utilized LS coupling scheme. Such calculations can potentially be coupled with the analytical treatment described by \citet{Judge+Casini+Paraschiv2021}, to formally overcome the same-ion degeneracy described by \citet{Dima+2020}. The effect is weaker than can be detected with current instrumentation.

The development of the CLEDB inversion \citep{par+judge2022} allows for inverting vector magnetic fields from Stokes $IQUV$ observations. As first shown by \citet{schad2024}, the 4 m large aperture and narrow field-of-view (FOV) Cryo-NIRSP \citep{fehlmann2016} spectrograph of the US National Science Foundation's Daniel K Inouye Solar Telescope \citep[DKIST][]{rimmele2020} is the first instrument realistically capable of measuring Stokes $V$ and enable Zeeman-diagnostics of the corona.  The planned 1.5 m aperture COronal Solar Magnetism Observatory \citep[COSMO][]{cosmo2022,cosmo2023} facility is provisioned to measure Stokes $V$ over a full-sun FOV,  while the CORSAIR balloon-borne spectropolarimeter \citep{samra2021corsair,samra2024} targets Stokes $V$ diagnostics as a pathfinder for future space-based instrumentation. 

The heliophysics community also employs a distinct method to infer coronal magnetic fields close to the plane-of-sky (POS) through the Sun's center, based on measuring magnetic properties of propagating Alfv\'{e}nic waves \citep{tomczyk2007} to derive POS projected magnetic field products (B$_{\text{POS}}$) as shown in \citet{morton+2016,yang2020+b,yang2024}. This method's feasibility is demonstrated observationally by stable high-cadence observations with a large FOV of the 0.2 m aperture Coronal Multi-channel Polarimeter \citep[CoMP;][]{tomczyk+2008} instrument operating at the Mauna Loa Solar Observatory (MLSO). An upgraded instrument, \citep[UCoMP;][]{2019shin.confE.131T} is currently in the commissioning phase at MLSO. The modest instrument aperture, along with the stringent photometric and polarimetric measurement requirements, makes it extremely difficult to unequivocally detect circular polarization signals in the corona, highlighting the need for large aperture instruments such as DKIST and COSMO. 

The diagnostic potential of Alfv\'{e}nic wave observations for magnetic field inference is far from fully exploited. \citet{Sharma2023}, along with studies employing novel Cryo-NIRSP observations \citep{morton2025a,morton2025b,molnar2026}, revealed that the corona is abundant with incompressible fluctuations most commonly interpreted as Alfv\'{e}nic waves, and have enabled further constraints on their origins. Such coronal waves are the crux of applying wave based magnetic field inferences, as shown by \citet{tomczyk2007,yang2020+,yang2024}. These studies reported differences of physical interpretation of wave propagation between open field regions and active coronal loops and hypothesized that the amplitudes of the Doppler velocity signal from these incompressible fluctuations are significantly reduced by LOS integration particularly in more LOS elongated regions like coronal hole and open field regions. Importantly, these works explicitly warn about bias in B$_{\text{POS}}$ inference. Therefore, further refinement of our understanding of coronal wave propagation is required before wave based measurements can be compared observationally with Zeeman measurements of coronal magnetic fields. 

%%Summary
This study aims to leverage information from diagnostics of Hanle saturated regime polarimetry with magnetic diagnostics of propagating Alfv\'{e}nic waves in order to bypass the need of measuring Stokes $V$. For demonstrating this framework, we compute and utilize a B$_{\text{POS}}$ diagnostic, analogous to one observationally inferred from wave propagation. We assume it is fully representative of the underlying coronal field, and show that it can in principle substitute a Zeeman-based diagnostic. This work will show that a novel ``$IQUD$'' (Stokes $IQU$ + Doppler oscillation based  B$_{\text{POS}}$) coronal seismology-assisted inversion can reproduce the information content of full-Stokes ``$IQUV$'' inversion \citep[introduced by][]{par+judge2022}, where the inversion performance is primarily limited by line-of-sight integration and SNR constraints, and to a lesser degree, by the accuracy of atomic data. These limitations define the regimes in which reliable coronal magnetic field inference is possible. The methodological reasoning and a synthetic benchmark dataset are presented in Sec.~\ref{sec:data}. A comparative statistical validation of the the novel $IQUD$ against $IQUV$ inversions, along with an expanded discussion on degeneracies and known assumptions and limitations, are presented in Sec.~\ref{sec:results}.  The results and broader implications of this approach towards furthering coronal magnetometry are summarized in Sec.~\ref{sec:concl}.

%----------------------------------------
\section{Methodological Description} \label{sec:data}
\subsection{Conceptual Framework}

We aim to investigate whether replacing the circular polarization Stokes $V$ with  magnetic diagnostics inferred from coronal seismology preserves the full magnetic and thermodynamic diagnostic power of the $IQUV$ inversion framework. Without additional assumptions, not enough information is encoded in just one M1 line to infer a complete vector (3D) magnetic field \citep{Plowman2014,Dima+2020,Judge+Casini+Paraschiv2021,par+judge2022}. The analytical and model comparison approaches are posed such that they take as input at least two emission lines, leading to at least eight wavelength-integrated Stokes observables. One-line inferences only have access to LOS projected fields, not the 3D field components. Several one-line approach example exist \citep{lin2000,Dima+2020,par+judge2022,schad2024}.

The implementation of the CLEDB inversion and software that we expand upon in this work is a ``single-point'' inversion algorithm for inferring magnetic fields using Stokes $IQUV$ observations of coronal M1 lines. The algorithm uses either the CLE or PyCELp forward synthesis codes to construct forward-modeled databases of Stokes $IQUV$ parameters for lines of interest, covering combinations of thermodynamic and magnetic configurations of coronal plasma. The size and complexity of such databases are drastically reduced by taking advantage of symmetries and by using wavelength-integrated line profiles. CLEDB then performs a straightforward least-squares sorting of database entries that best fit an input two-line observation. 

In the $IQUV$ inversion, the inputs are wavelength-integrated measurements of Stokes $IQUV$ in both \ion{Fe}{13} lines and the POS line-ratio electron density ($n_e$) $(IQUV_{1074}\,,IQUV_{1079} \,,n_e)$, while for the $IQUD$ setups, the input variable for Stokes $V$ is replaced with an B$_{\text{POS}}$ estimation, as resulting from coronal wave-tracking analysis $(IQU_{1074}\,,IQU_{1079} \,, n_e\,, \text{B}_{\text{POS}}$). A degenerate set of solutions for the magnetic field unit vector $\hat{b}$ is returned, resembling inversions applied to solar photospheric magnetic fields. As Stokes $V$ $\varpropto d I / d\lambda$, we account for the line width when calculating integrated quantities such that the weak-field Zeeman response can be approximated as linear with respect to the field strength $B$. Thus, we are able to algebraically recover $B$, through the ratio of observed to forward-modeled Stokes $V$ \citep{par+judge2022}.

In the saturated regime of the Hanle effect, the Stokes vector of an M1 emission line can be written as follows,

\begin{eqnarray}
    I \quad \varpropto &\quad  I_0\; [\: 1 \; + \, \frac{1}{2\sqrt{2}}\: D_{\scriptscriptstyle JJ0}\, & \sigma^2_{0}(\vartheta_{B})_{\scriptscriptstyle J} \qquad\, (3\cos^2 \Theta_{B} - 1)\; ]  \label{eqn:line-1} \\
    Q \quad \varpropto &\quad\, I_0 \hspace{3.0em} \frac{3}{2\sqrt{2}}\; D_{\scriptscriptstyle JJ0}\; & \sigma^2_{0}(\vartheta_{B})_{\scriptscriptstyle J} \qquad\quad\, \sin^2 \Theta_{B} \;\, \cos 2\Phi_{B} \label{eqn:line-2} \\
    U \quad \varpropto & \; -\, I_0 \hspace{3.0em} \frac{3}{2\sqrt{2}}\; D_{\scriptscriptstyle JJ0}\; & \sigma^2_{0}(\vartheta_B)_{\scriptscriptstyle J} \qquad\quad\, \sin^2 \Theta_B \;\, \sin 2\Phi_B \label{eqn:line-3} \\
    V \quad \varpropto & -\, I_0\; [\: \bar{g}_{\scriptscriptstyle JJ0}\;\; +\;\; E_{\scriptscriptstyle JJ0} \,& \sigma^2_{0}(\vartheta_B)_{\scriptscriptstyle J}\,] \;\;\;\, \nu_{\scriptscriptstyle L} \;\,\,\cos\;\, \Theta_B\,. \qquad\qquad\quad~, \label{eqn:line-4} 
    %V \quad \varpropto & \, -\, C_{\scriptscriptstyle JJ_0}\; [\: \bar{g}_{\scriptscriptstyle JJ0}\, +\quad E_{\scriptscriptstyle JJ0} \,& \sigma^2_{0}(\vartheta_B)_{\scriptscriptstyle J}\,] \; \omega_{\scriptscriptstyle B} \,\cos\: \Theta_B \qquad\qquad\quad~~ \label{eqn:line-4} \tfrac{\lambda^2}{c}
\end{eqnarray}

These equations represent a formulation of the established spectral line formation formalism of \citet{Casini+Judge1999} as derived in Eqs.~17--20 in \citealt{judge2007} and Eqs.~1--4 in \citealt{Dima+2020}, emphasizing the dependence of the emergent Stokes $IQUV$ profiles on the local magnetic properties at each observed location. $I_0$ is a constant representing the unpolarized line emission integrated over a homogeneous path length in the optically-thin corona. The angles $\Phi_B$ and $\Theta_B$ describe the magnetic-field orientation on the POS and along the LOS respectively. This formulation implicitly assumes that the coronal emission originates from a geometrically thin, dominant structure along the LOS, with respect to the full integrating path length. The $D_{\scriptscriptstyle JJ_0}$,  $E_{\scriptscriptstyle JJ_0}$, and $\bar{g}_{\scriptscriptstyle JJ0}$ represent atomic transition parameters for each considered M1 line that are not directly related to the magnetic orientation. $\sigma^2_0$ is the \emph{atomic alignment factor}, which quantifies the population imbalance among the magnetic sublevels of the atomic level. It arises from the anisotropy of the photospheric radiation field, which is height-dependent in the solar atmosphere. This quantity can be modified by collisional processes governed by $n_e$ and electron temperature ($T_e$), as well as by the inclination of the magnetic field with respect to the local solar vertical ($\vartheta_B$). The alignment can take both positive and negative values depending on the interplay of these mechanisms. Because we are neglecting the contribution of the transverse Zeeman effect, the  Stokes $Q$ and $U$ are practically proportional to $\sigma^2_0$. When searching for a best-fit solution through discretized positions along the LOS, CLEDB exploits the $\sigma^2_0$ dependence on $\vartheta_B$ to find the ``true height'' of the dominant emission \citep{Judge+Casini+Paraschiv2021}. The coupling between magnetic geometry and emission height plays a key role in constraining the inversion, and ultimately, the retrieved coronal magnetic structure. The relevant geometrical definitions are illustrated in Fig.~5 of \citet{Casini+Judge1999} and explicitly summarized in Table~1 of \citet{par+judge2022}.

Sensitivity to $B$ appears only in Eq.~\ref{eqn:line-4}, through the Larmor frequency $\nu_{\scriptscriptstyle L}$. Because Eqs.~\ref{eqn:line-1}--\ref{eqn:line-3} are derived in the saturated Hanle regime, Stokes $I$, $Q$, and $U$ depend only on the magnetic field direction ($\hat{b}$). The squared geometrical dependence for both $\Theta_B$ and $\Phi_B$ angles introduces additional degeneracies, revealing that one-line observations of Stokes $IQU$ are insufficient to solve for the 3D magnetic field vector. 

To make the problem tractable, suppose we select a two-line combination to observe $IQU$ quasi-simultaneously, with a current generation instrument such as Cryo-NIRSP or UCoMP. The \ion{Fe}{13} 1074.7 nm and 1079.8 nm line pair is currently the most feasible choice in order to avoid requiring to constrain abundance ratios of different atoms. The resulting six Stokes observables can be used to return a set of 4-times degenerate solutions corresponding to $\hat{b}$. In such a setup, any independent knowledge of $B$ or its projections $B_{\text{LOS}}$ or $B_{\text{POS}}$ can be added to a $IQU$-only inversion, to further constrain the retrieved forward-modeled solution. We draw upon the work of \citet{yang2020+,yang2020+b} which use Doppler oscillations to infer the properties of propagating Alfv\'{e}nic waves to constrain coronal magnetic structures. Namely, we adopt the following simplification for the dispersion relation for propagating kink waves resulting from tracking Doppler oscillations following the form given by \citet{yang2020+b}:
\begin{equation}
    c_k=\dfrac{B}{\sqrt{\mu_0\:\langle \rho \rangle}},
    \label{eqn:phase1}
\end{equation}
where $\mu_0$ is the vacuum magnetic permeability, $c_k$ represents the wave phase speed, and $\langle \rho \rangle$ is the average plasma mass density. Following \citet{tian2012}, $\langle \rho \rangle \approx 1.2\,n_e\,m_p$,  where $m_p$ is the proton mass and $n_e$ is derived via \ion{Fe}{13} intensity line-ratios using an atomic database \citep[e.g. CHIANTI;][]{dere1997,delzanna2021}. 

Directly implementing Eq.~\ref{eqn:phase1} will raise an important assumption. When interpreting the phase speed, we associate it with a wave propagating radially in the POS. In turn, this assumption implies that a CLEDB inversion should only retrieve database solutions with inclinations very close to the POS; i.e. with $\sin \Theta_B\approx 1$, effectively removing the LOS position as a free parameter. In consequence, the degeneracy discussed in \citet{Dima+2020} and \citet{Judge+Casini+Paraschiv2021} would be reintroduced.

Alternatively, we adopt the interpretation of \citet{yang2020+b} that the measured phase speed corresponds to the POS component of the true phase speed. The inferred magnetic field strength likewise corresponds to its POS component. Suppose we detect a wave traveling at an inclined angle with respect to the POS. For an observed $c_k$ wave speed, this implies that instead of relating directly to the $B$, we are projecting $B$ on the POS plane as

\begin{equation}
    c_{km}=\dfrac{B\, \sin \Theta_B}{\sqrt{\mu_0\:\langle \rho \rangle}}.
    \label{eqn:phase2}
\end{equation}

This relaxed constraint enables LOS position retrieval with CLEDB by allowing magnetic-field orientations that are not strictly confined to the POS. In this framework, the physical true height of the dominant structure is inferred by exploiting the sensitivity of $\sigma^2_0$  discussed above. Consequently, this setup will allow for finding database combinations of $\Theta_B$, $\sigma^2_0$, and $B$ that are compatible with an observed $c_{km}$, theoretically allowing for observing magnetic structures in front or behind the POS, or even structures that cross through the POS. $B \sin \Theta_B$ can theoretically be applied to further refine an $IQUD$ inverted set of initial 4-times degenerate solutions, to constrain an inversion to a set of 2-times degenerate solutions, as resulting from a $IQUV$ inversion. Detailed discussions follow in Sec.~\ref{ssec:degen}.

%----------------------------------------
\subsection{Forward Modeled Synthetic Observations Dataset }

\begin{deluxetable}{lccc}
\tablenum{1}
\tablecaption{\fontsize{10}{12}\selectfont Synthetic Observation Parameter Sample Ranges \label{tab:grid}}
%\tablewidth{0pt}
\tablehead{
\colhead{Parameter} & 
\colhead{Units} & 
\colhead{No. Samples} & 
\colhead{Range}
}
\startdata
$y\;\:$ (projected height)  & $R_{\odot}$     & $\,\; n_y = 20$       & 1.01 --- 1.90  \\
$n_e$ (electron density) & \;\;\;cm$^{-3}$ & $n_{n_e} = 62$        & $\;\,10^{7.00}$ --- $10^{10.05}$ \\
$x\;\:$ (LOS position)   & $R_{\odot}$     & $\,\;n_x = 21$        & $\!\!\!\!\!-0.5$ --- 0.5  \\
$\phi \;\:$ (azimuthal angle) & rad             & $\;\;\;n_\phi=180$ & $\;\;0$ --- $2\pi$  \\
$\theta \;\;$ (polar angle)   & rad             & $\,\;n_\theta=90$  & $\;\,0$ --- $\;\pi$ \\
\enddata
\tablecomments{\fontsize{10}{12}\selectfont The parameters  are sampled using the same discretization adopted in the CLEDB database. $\phi$ and $\theta$ follow the definitions in  \citet[Tab. 1]{par+judge2022}, with a correction to the definition of $\phi$ given by $\phi = \operatorname{arctan}
\frac{\hat{x}\cdot(\hat{b}\times\hat{z})}{
\hat{x}\cdot\hat{b}}.$}
\end{deluxetable}

Aiming to study the fundamental application of both  $IQUV$ and $IQUD$ inversion setups with respect to a broad range of plasma and magnetic geometries applicable to coronal observations, we create a dataset of synthetic coronal observations using PyCELp. Tab.~\ref{tab:grid} shows the ranges of observed projected height ($y$), $n_e$, LOS position ($x$), along with the $\phi$, and $\theta$ magnetic angles. In the forward synthesis of this dataset, as well as in the CLEDB databases, we assume a fixed peak ionization temperature of $\log T_e=6.22$, corresponding to \ion{Fe}{13}, and a coronal magnetic field strength of $B = 1G$. Temperature variations are not considered in the present analysis, as the thermal state of the emitting plasma cannot be constrained well enough to support a more general treatment. Under this peak-temperature assumption, a $1G$ field produces a very weak Stokes $V$ signal, on the order of $\sim10^{-5}$ relative to Stokes $I$, affecting detectability in low-SNR observations.

\begin{figure}
\centering
\includegraphics[width=\linewidth]{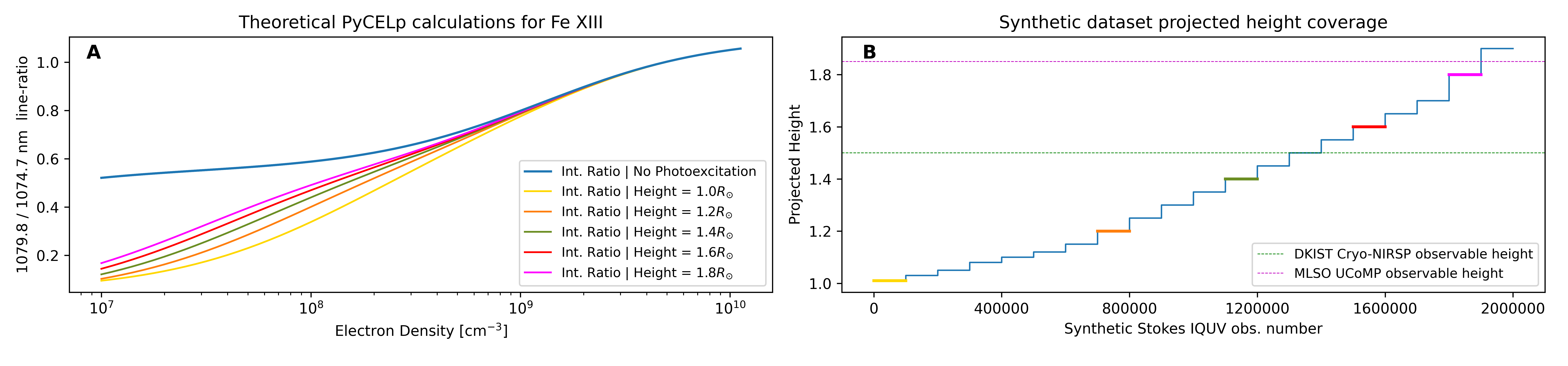}
\caption{\fontsize{10}{12}\selectfont Overview of the physical diagnostics with respect to the height sampling of the forward-modeled Stokes $IQUV$ synthetic dataset. (A) Theoretical $n_e$ dependence of the \ion{Fe}{13} IR line-ratio including photoexcitation effects, shown for five different emission height subsets, measured from disk center in units of $R_{\odot}$. (B) The projected height discretization covered in the dataset. The five predefined subsets of $10^5$ observations corresponding to the line-ratio calculations presented in panel (A) are highlighted.}
\label{fig:obsparams}
\end{figure}

A statistically significant dataset of $2\cdot10^6$ synthetic Stokes $IQUV$ observations was generated for each of the two IR \ion{Fe}{13} emission lines, as follows: The Latin Hypercube Sampling (LHS) technique \citep{Iman1981} was used to efficiently sample plasma conditions and geometries, while maintaining uniform coverage of the parameter ranges. Fig.~\ref{fig:obsparams} (A) shows six representative theoretical calculations of \ion{Fe}{13} densities derived via line-ratios, highlighting the line pair formation regime that entails significant contributions in both photoexcitation and collisional excitation that are acting differentially across observable heights. Because of the mixing of these excitation regimes, an uncertainty in diagnostics accuracy when using this line pair exists \citep{delzanna2023_airspec,delzanna2025}. To best mitigate this effect, LHS was used to extract samples of $n_e$, $x$, $\phi$, and $\theta$ spanning the ranges listed in Tab.~\ref{tab:grid}, for 20 representative $y$ values, because projected heights can be directly constrained. Fig.~\ref{fig:obsparams} (B) shows how the resulting dataset is divided into $y$ subsets of $10^5$ observations.

The LHS samples are subsequently mapped onto the discrete sampling grid shown in Tab.~\ref{tab:grid} in order to reproduce the CLEDB database configuration and discretization scheme used in CLEDB~V0.7.0. LHS samples can collapse onto identical grid points after discretization, introducing redundancy in the mapped parameter space. We quantified this effect through numerical tests, finding a small loss of unique realizations per $y$ value. We therefore increased the number of LHS samples by 0.1\%, which is sufficient to ensure $10^5$ unique parameter combinations per $y$ after grid mapping. The sampled and grid-mapped parameters are then forward-synthesized into Stokes $IQUV$  observations. Consequently, the synthetic Stokes $IQUV$ are generated on the same parameter grid as the CLEDB database. This design ensures that near-exact database solutions exist for the synthetic observation dataset, and that the corresponding ground-truth solution is known a priori. When discussing inversion results, we therefore define a \emph{match} as any database solution that reproduces the input plasma parameters and magnetic geometry when accounting for inversion degeneracies. When assuming ``noise-free'' observations, this setup effectively isolates intrinsic inversion degeneracies, while avoiding ambiguities introduced by database discretization.

To run the $IQUD$ inversion setup we require B$_{\text{POS}}$ and compute it from the magnetic geometries used in forward calculating the dataset entries. This serves as a proxy for a waves inferred B$_{\text{POS}}$.

Furthermore, as a first step toward assessing the applicability of the inversion framework to real observations, we apply randomized multiplicative perturbations to the sets of two-line Stokes $IQUV$ amplitudes  as $IQUVIQUV_{pert}$ = $IQUVIQUV_{true}(1+\epsilon)$, where $\epsilon$ denotes zero-mean relative uncertainty realizations applied independently to each individual Stokes component in a set. We consider five relative uncertainty levels with $\epsilon$ drawn from distributions corresponding to ranges of $10^{-5}$, $10^{-4}$, $10^{-3}$, $10^{-2}$, and $5\cdot10^{-2}$ of each respective forward-calculated Stokes $I$ component. The range samples a broad set of conditions, where $10^{-5}$ corresponds to an excellent polarization measurement, $10^{-4}$ and $10^{-3}$ correspond to a generally achievable measurements, while $10^{-2}$ and  $5\cdot 10^{-2}$, corresponds to measurements in adverse conditions. The perturbation distributions from which $\epsilon$ are drawn follow a Gaussian model with a 3$\sigma$ confidence, assuring  $\sim99.7\%$ coverage probability that $\epsilon \in [-l,l]$, where \emph{l} denotes the five relative thresholds. These perturbations are applied to the entire dataset of $2\cdot10^6$ polarimetric observations. The resulting noise-free and the five perturbed datasets are inverted separately in both $IQUV$ and $IQUD$ configurations. 

This perturbation experiment does not directly translate to observational constraints. Our forward synthesized profiles do not explicitly model varying coronal backgrounds up to background-limited regimes, that will contribute in a non-multiplicative way to an observation, influencing the SNR as shown in Sec. 2 of \citet{penn2004}. In addition, estimating a realistically constrained observational SNR with both additive and multiplicative terms depends on multiple situational factors like: the combinations of spatial, spectral, and temporal resolution of an observation, short-time varying observing conditions, background subtraction residuals, etc. Developing a model fully representative of realistic observational conditions was outside the scope of this work, though recent efforts to propagate calibration and instrumental uncertainties into Stokes and magnetic-field reconstructions \citep{Hsu2026} point toward a viable path for such future work.

%----------------------------------------
\section{Analysis and Results} \label{sec:results}
\subsection{Statistically Validating CLEDB Using Synthetic Observations}\label{ssec:stat}

To assess robustness, we compare the $IQUV$ and $IQUD$ inversion outputs against the ground-truth physics across the full dataset. Noise-free and uncertainty-perturbed observations are treated as separate benchmarking cases, allowing the effects of measurement noise to be disentangled from those of intrinsic degeneracies. The interplay between line formation regimes and recovered physical parameters is examined.

The  ``\% Matched Obs.'' column  of Tab.~\ref{tab:uncertainty} presents the main inversion outputs for both inversion setups. When discussing the entire dataset, we find that in the noise-free case we match $\sim$99\% of the input observations. We find no significant difference between the $IQUV$ and $IQUD$ inversion results, aside from the degeneracy multiplicity. The $\sim$1\% of incorrect matches arise from a distinct class of geometrical degeneracies that are further discussed in Sec.~\ref{ssec:degen}. In the five perturbed cases we find that the physics and magnetic geometry can be recovered with minimal loss (matches $>96\%$) up to perturbation levels of $10^{-3}$. Even at the $10^{-2}$ level, a majority of input observations are correctly matched ($>77\%$).

%% R3 version with LHS - no deg 
\begin{deluxetable}{l c c c c c}
\tablenum{2}
\tablecaption{\fontsize{10}{12}\selectfont Percentage of synthetic observations that are matched by the CLEDB inversion in a noise-free case and five different perturbation levels. \label{tab:uncertainty}}
\tablewidth{0pt}
\tablehead{
\colhead{}  & \colhead{Uncertainty Level} & \multicolumn{2}{c}{\% Matched Obs.\tablenotemark{1,2}}  & \multicolumn{2}{c}{\% Geometrical Match Obs.\tablenotemark{1,3}} \\
\colhead{} & \colhead{} & \colhead{\scriptsize $IQUV$ $|$ $IQUD$} &  \colhead{\scriptsize $n_e < 7.15
\;|\;n_e > 9.90$}    & \colhead{\scriptsize $IQUV$ $|$ $IQUD$}  & \colhead{\scriptsize Err. width: $1\sigma|2\sigma$} 
}
\startdata
\multirow{5}{*}{\rotatebox{90}{\scriptsize All Heights}} 
\multirow{5}{*}{\rotatebox{90}{\scriptsize 2000000 obs.}} 
    & $5\cdot10^{-2}\quad$ & 30.0 $|$ 30.1 & +7.5 $|$ -5.7  & 52.3 $|$ 52.4 & $\:20^\circ~|~221^\circ$\\
    & $10^{-2}$ & 76.8 $|$ 77.0  & +7.4 $|$ -8.9  & 89.6 $|$ 89.9 & $~0^\circ~|~31^\circ$\\
    & $10^{-3}$ & 96.2 $|$ 96.7 & +0.3 $|$ -1.0  & 98.4 $|$ 98.8 & $0^\circ~|~0^\circ$\\
    & $10^{-4}$ & 98.1 $|$ 98.6  & +0.0 $|$ -0.0 & 98.5 $|$ 98.9 & $0^\circ~|~0^\circ$\\
    & $10^{-5}$ & 98.4 $|$ 98.9  & +0.0 $|$ -0.0 & 98.6 $|$ 98.9 & $0^\circ~|~0^\circ$\\
    & noise-free     & 98.9 $|$ 98.9  & $\;\,$+0.0 $|$ +0.0  & 98.9 $|$ 98.9 & $0^\circ~|~0^\circ$\\
    %&\color{red} pure-LHS     & $|$   &  $|$   &\color{red} 84.78 $|$\color{red} 85.03 & {\color{red} $~0^\circ~|~33^\circ$}\\
    %&\color{red} pure-LHS  $<10^\circ$   & $|$   &  $|$   &\color{red} 94.76 $|$ \color{red} 94.92 & {\color{red}$~0^\circ~|~11^\circ$}\\
    \tableline
\multirow{3}{*}{\rotatebox{90}{\scriptsize h$\sim$1.01$R_{\odot}$}} 
\multirow{3}{*}{\rotatebox{90}{\tiny 100000 obs.}} 
    & $10^{-2}$ & 86.4 $|$ 86.8 & +1.3 $|$ -2.1  & 93.0 $|$ 93.3 & ~$0^\circ~|~17^\circ$\\
    & $10^{-3}$ & 97.1 $|$ 97.6 & +0.1 $|$ -0.5  & 98.4 $|$ 98.9 & $0^\circ~|~0^\circ$\\
    & noise-free     & 98.9 $|$ 98.9 & $\;\,$+0.1 $|$ +0.2  & 98.9 $|$ 98.9 & $0^\circ~|~0^\circ$\\
    \tableline
\multirow{3}{*}{\rotatebox{90}{\scriptsize h$\sim$1.20$R_{\odot}$}} 
\multirow{3}{*}{\rotatebox{90}{\tiny 100000 obs.}} 
    & $10^{-2}$  & 78.6 $|$ 78.9 & $\;\,$+8.6 $|$ -12.9 & 90.4 $|$ 90.7 & ~$0^\circ~|~27^\circ$\\
    & $10^{-3}$  & 97.0 $|$ 97.4 & +0.2 $|$ -0.9 & 98.4 $|$ 98.8 & $0^\circ~|~0^\circ$\\
    & noise-free     & 98.9 $|$ 98.9 & $\;\,$+0.2 $|$ +0.1  & 98.9 $|$ 98.9 & $0^\circ~|~0^\circ$\\
    \tableline
\multirow{3}{*}{\rotatebox{90}{\scriptsize h$\sim$1.40$R_{\odot}$}} 
\multirow{3}{*}{\rotatebox{90}{\tiny 100000 obs.}} 
    & $10^{-2}$  & 75.0 $|$ 75.2 & $\;\,$+8.9 $|$ -10.2  & 88.7 $|$ 88.9 & ~$0^\circ~|~34^\circ$\\
    & $10^{-3}$  & 96.0 $|$ 96.5 & +0.1 $|$ -1.1  & 98.4 $|$ 98.8 & $0^\circ~|~0^\circ$\\
    & noise-free     & 98.9 $|$ 98.9 & $\;\;\;$-0.1 $|$ +0.1 & 98.9 $|$ 98.9 & $0^\circ~|~0^\circ$\\
    \tableline
\multirow{3}{*}{\rotatebox{90}{\scriptsize h$\sim$1.60$R_{\odot}$}}
\multirow{3}{*}{\rotatebox{90}{\tiny 100000 obs.}} 
    & $10^{-2}$  & 72.2 $|$ 72.4 & +9.8 $|$ -8.2 & 87.8 $|$ 88.1 & ~$0^\circ~|~36^\circ$\\
    & $10^{-3}$  & 95.8 $|$ 96.2 & +0.6 $|$ -1.6 & 98.4 $|$ 98.8 & $ 0^\circ~|~0^\circ$\\
    & noise-free     & 98.9 $|$ 98.9 & $\;\,$-0.0 $|$ -0.0  & 98.9 $|$ 98.9 & $0^\circ~|~0^\circ$\\
    \tableline    
\multirow{3}{*}{\rotatebox{90}{\scriptsize h$\sim$1.80$R_{\odot}$}}
\multirow{3}{*}{\rotatebox{90}{\tiny 100000 obs.}} 
    & $10^{-2}$  & 69.1 $|$ 69.3 & +9.2 $|$ -6.7 & 86.9 $|$ 87.1 & ~$0^\circ~|~39^\circ$\\
    & $10^{-3}$  & 94.7 $|$ 95.1 & +0.8 $|$ -1.8 & 98.3 $|$ 98.7 & $ 0^\circ~|~0^\circ$\\
    & noise-free     & 98.9 $|$ 98.9 & $\;\;\;$-0.0 $|$ +0.0 & 98.9 $|$ 98.9 & $0^\circ~|~0^\circ$\\
    \tableline 
\enddata
\tablenotetext{1}{\fontsize{10}{12}\selectfont For the $IQUV$ inversion, a match is sought from the first two $\chi^2 $sorted inversion solutions. For the $IQUD$ inversion a match is searched for inside the first four $\chi^2$ sorted inversion solutions.}
\tablenotetext{2}{\fontsize{10}{12}\selectfont Variation in solution matching sensitivity with respect to POS $n_e$, when considering the lower and upper 5 percentiles of sampled densities in each of the observation subsets ($n_e<7.15$ and $n_e > 9.90$). The third column reports the percentage variation in matches relative to the reference values listed in the second column. A single metric is reported, as no significant difference was found between the $IQUV$ and $IQUD$ setups.}
\tablenotetext{3}{\fontsize{10}{12}\selectfont Solutions with geometries consistent with the ground-truth within one CLEDB database discretization step. The reported distribution of angular widths corresponds to the differences between the observed and best-matching solutions in $\phi$ and $\theta$ (degrees), evaluated at the 68\% and 95\% percentiles of the observation sample, including exact matches.}
\tablecomments{\fontsize{10}{12}\selectfont Results are obtained using CLEDB~V0.7.0. At the cost of trading accuracy by lowering the database resolution, the matches presented in this table can be significantly improved.}
\end{deluxetable}

A key question is how many of the incorrect matches remain close to the ground-truth geometry. We introduce an additional metric where a solution is considered ``geometrically matched'' when the retrieved $\phi$ and $\theta$ lie within $\pm2^\circ$, and $x$ is within $\pm0.05R_\odot$ of the ground-truth. In other words the retrieved solution lies within $\pm1$ discretization step from the ground-truth ($\pm x\times\phi\times\theta$).  Allowing for this geometrical relaxation increases the number of acceptable inversion solutions in both the $IQUV$ and $IQUD$ cases, as expected from adopting a less restrictive matching criterion. Even for adverse perturbations of $10^{-2}$ and $5\cdot10^{-2}$, more than 89\% and 52\% of solutions, respectively, remain close to the ground-truth geometry. The last column of Tab.~\ref{tab:uncertainty} presents the $1\sigma$ (68\%) and $2\sigma$ (95\%) angular error widths of the retrieved $\phi$ and $\theta$ values across the dataset. The $1\sigma$ threshold width remains below $20^\circ$ for the $5\cdot10^{-2}$ perturbations, indicating that qualitative magnetic orientations might still be recoverable, even in the case adverse observing conditions.

Due to the complex IR \ion{Fe}{13} line formation mechanisms, we carefully examine the interplay between the physical parameters involved in generating the Stokes $IQUV$ profiles, by scrutinizing the height dependent reliability of both $IQUV$ and $IQUD$ inversions. We use the five representative projected heights at $1.01R_{\odot}$, $1.20R_{\odot}$, $1.40R_{\odot}$, $1.60R_{\odot}$, $1.80R_{\odot}$, shown in Fig.~\ref{fig:obsparams} (B), where the primary line-ratio density diagnostic is easily  distinguishable as shown in Fig.~\ref{fig:obsparams} (A). These five representative subsets are then inverted noise-free and with the $10^{-3}$ and $10^{-2}$ perturbation levels that were found above to be most relevant. The inversion results are appended to Tab.~\ref{tab:uncertainty}. The negligible difference between $IQUV$ and $IQUD$ inversion setups persists across these observational subsets.  As the sampled projected height increases, the coronal signal decreases as expected, but the relative contributions of photoexcitation versus collisional excitation also change, modifying the resulting matches. When comparing the $1.01R_{\odot}$ with the $1.80R_{\odot}$ subsets, the fraction of incorrect matches increases from negligible in the noise-free case, to $\sim2\%$ for the $10^{-3}$ perturbation, and up to $\sim17\%$ for the adverse $10^{-2}$ perturbed observations. This indicates that the evolving balance between the two excitation mechanisms in the coronal plasma significantly influences inversion performance in addition to uncertainties.

\begin{figure}
\centering
\includegraphics[width=\linewidth]{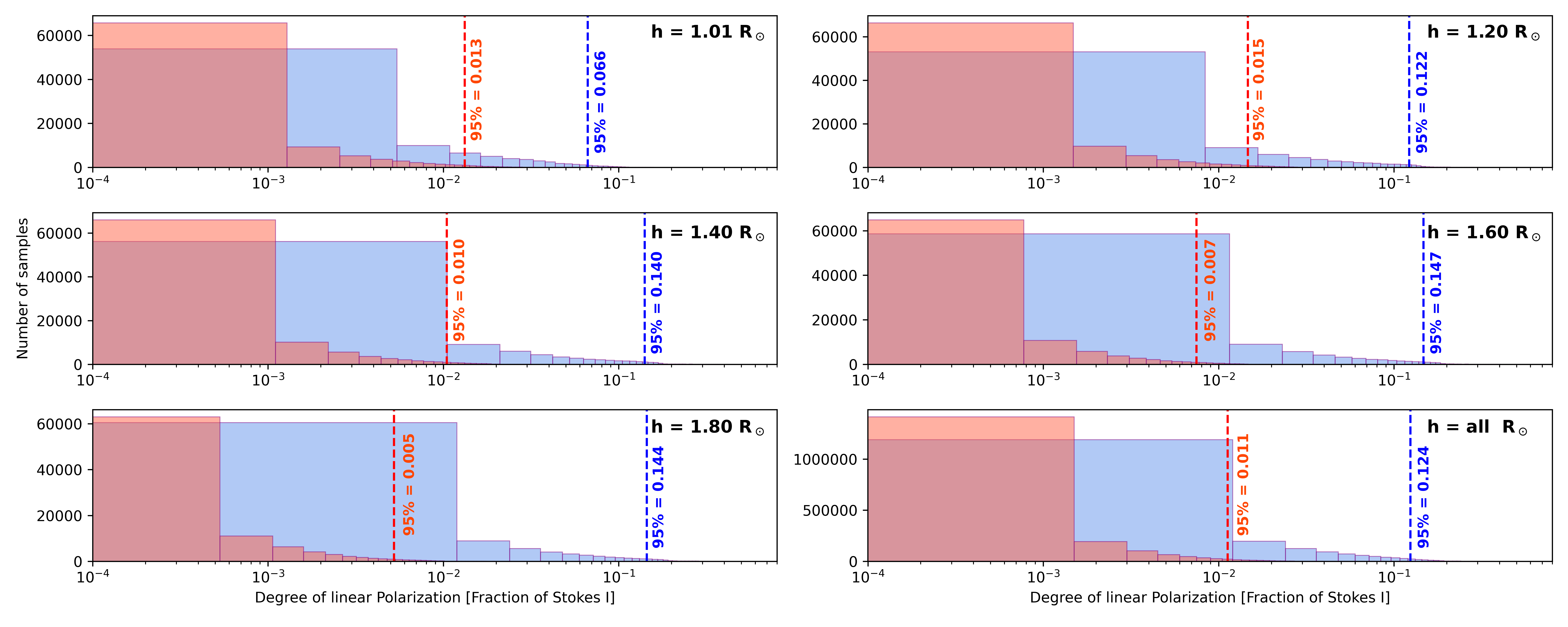}
\caption{\fontsize{10}{12}\selectfont Distributions of the degree of linear polarization for the entire dataset and five subsets shown in Fig.~\ref{fig:obsparams} (B). Histogram colors denote the \ion{Fe}{13} 1074.7 nm (blue) and 1079.8 nm (red) lines. The 95th percentile values of the distributions across the sampled height range are indicated using the same color scheme.}
\label{fig:linpol}
\end{figure}

This observed loss of accuracy is caused by a combination of effects. Theoretically, the anisotropy of the radiation field will increase with height, and increase the degree of linear polarization. At the same time, the radiation field gets weaker with increasing height, and collisions start to dominate the line formation. As opposed to photoexcitation, polarized line formation in the presence of collisions is not a height dependent process, but is directly proportional to the amount of plasma (i.e. $n_e$) contributing to an observation. In a coronal case, the $n_e$ is generally expected to decrease as a function of height, leading to decreased collisions. In addition, the IR \ion{Fe}{13} lines have different maximum degrees of polarizability, adding to these sensitivity issues. The superposition of these effects are shown in Fig.~\ref{fig:linpol} through distributions of the degree of linear polarization of dataset entries. The 95 percentiles of the distributions show that for the 1074.7 nm line (blue), the degree of linear polarization increases up to heights of 1.60 R$_{\odot}$, but afterward, the line formation becomes dominated by collisional excitation. In the case of the 1079.8 nm line (red), this change in the formation regime occurs at significantly lower heights at $\sim$1.20 R$_{\odot}$. Thus, we find that for the perturbed observations, the reduced degree of linear polarization at larger heights leads to a measurable decrease in inversion reliability.

Although in an idealized steady-state corona we expect a general drop in density, this can not be assumed to be a universal case. Thus, Tab.~\ref{tab:uncertainty} also presents the dependence of matched solutions in both $IQUV$ and $IQUD$ cases on outlying $n_e$ values. We recalculated the matches considering only $n_e$ values in the first and last 5 percentile of the density range of Tab.~\ref{tab:grid}. No significant change in the inversion matches manifests up to $10^{-3}$ perturbations. At low $y$ of $R_{\odot}<1.20R$ the effects are only modest even with significantly $10^{-2}$ perturbed observations. At higher $y$, both $n_e$ subsets start to significantly influence inversions, where low $n_e$ increases matches and high $n_e$ decreases matches. Although this ``density sensitivity effect'' with respect to SNR levels is found to be beneficial in low $y$ and low $n_e$ cases, we find that collisional depolarization is amplified in regions of high $n_e$, even at higher heights where the photoexcitation would otherwise be low. The density sensitivity, combined with the height-dependent decrease in linear polarization fraction discussed above, can adversely impact inversions, at higher heights, higher densities, or both. 

Overall, we find that the Stokes observables emerge from the complexly-coupled effects of the plasma physical state, observational geometries, and magnetic geometries. Within the probed parameter ranges and heights of 1-2~$R_{\odot}$ from disk center, CLEDB inversions robustly recover the physical parameters and magnetic geometry from the \ion{Fe}{13} IR line pair in both $IQUV$ and $IQUD$ inversion modes. Furthermore, CLEDB correctly retrieved LOS position information, despite the input observations providing only $y$ and POS $n_e$ constraints. This supports the conclusions of \citet{Dima+2020} and \citet{Judge+Casini+Paraschiv2021} that additional physical constraints, such as the LOS dependence of the $\sigma^2_0$ alignment, can alleviate degeneracies present in analytical inversion schemes \citep{Plowman2014,Dima+2020}. Such analytical schemes can not include model excitation mechanisms and density dependencies in sufficient detail to recover LOS position information.

%----------------------------------------
\subsection{Intrinsic Degeneracies}
\label{ssec:degen}

\begin{figure}
\centering
%\rule{0.45\linewidth}{0.45\linewidth}
%\includegraphics[width=\linewidth]{figures/fig3_fovs_r1.png}
\includegraphics[width=\linewidth]{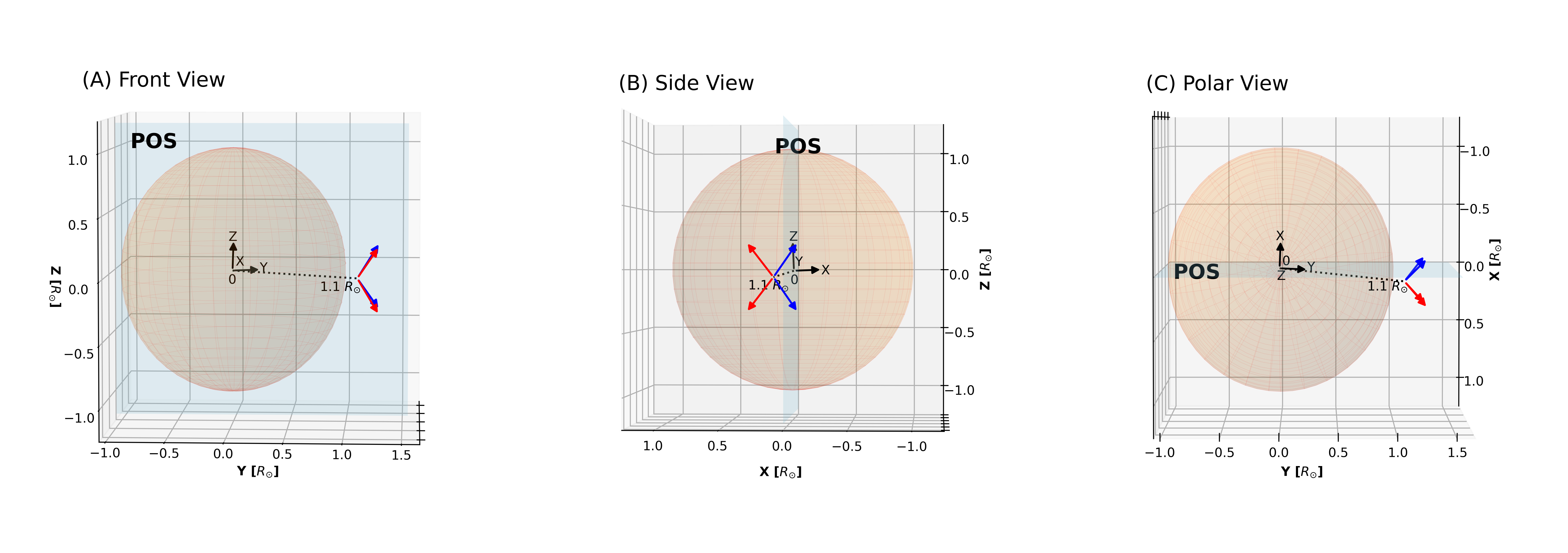}
\caption{\fontsize{10}{12}\selectfont Example of degeneracies in $IQUV$ and $IQUD$ inversions. One 4-times degenerate magnetic solution is represented, where the blue and red arrows show degeneracies with respect to the $x$ and $z$ planes. Three representative viewports are selected. (A) A POS observer projection, analogous to a ground-based observation. (B) The same projection but with the equatorial plane rotated by $90^\circ$. (C) Polar north looking down projection perpendicular to the POS.}
\label{fig:degeneracy}
\end{figure}

It is imperative to discuss a number of sources of degeneracy found when performing this analysis in both $IQUV$ and $IQUD$ inversion setups. Tab.~\ref{tab:uncertainty} shows that even when designing our input observation dataset configuration to reproduce CLEDB database entries, a subset of inversion solutions in the noise-free case exhibit additional degeneracies that are not directly solvable by our approach: 

A small subset of solutions corresponds to configurations in which the magnetic field has a negligible LOS component, such that Stokes $V$ vanishes and the magnetic field strength cannot be constrained within the $IQUV$ inversion setup. These cases occur when the field lies entirely in the POS or in any plane parallel to it within the allowed LOS positions (transverse to $x$ in Fig.~\ref{fig:degeneracy}). In such configurations the LOS position is only weakly constrained by the $\sigma_0^2$ of the two sets of Stokes observations and becomes degenerate.  The $\phi$ and $\theta$ orientations might be correctly inferred through the linear polarization azimuth provided $\theta\ne 0$ \citep{judge2007,judge2013}, even though the inversion cannot properly scale the magnetic field strength, and therefore all retrieved $B_x,\;B_y, \text{and } B_z$ are set to undefined by default. An important distinction emerges for the $IQUD$ configuration, where the primary constraint is the POS projection of the magnetic field. In this case, the field strength and the $B_x$, $B_y$, and $B_z$ components can still be recovered reliably. The analogous situation also applies, where if the field is longitudinally oriented with next to none POS projection, the $IQUD$ inversion will lead to undetermined $B$, and the only option for retrieving the field strength becomes an $IQUV$ inversion. Taken together, these two limiting cases indicate that $IQUV$ and $IQUD$ are complementary across the range of $\Theta_B$: $IQUD$ constrains the field best when it is closer to the POS and degrades as the field becomes more longitudinal ($\sin \Theta_B\rightarrow\pm1$), while $IQUV$ shows the opposite trend ($\sin \Theta_B\rightarrow0$), such that the applicability of either of the two methods will overlap over some intermediate range of $\Theta_B$. We note that the practical retrieval of either B$_{\text{POS}}$ through coronal seismology or B through Stokes $V$ will each preferentially respond to certain magnetic orientations.

The $\sigma^2_0$ atomic alignment, is functionally dependent on $(3\cos^2\vartheta_B-1)$ leading to it vanishing at $\vartheta_B = 54.74^\circ$ or $125.26^\circ$. This is known as the  Van Vleck Effect \citep{vanvleck1925}. In this case, the observation becomes insensitive to the radiation field anisotropy ($\sigma^2_0 \to 0$). As can be seen from Eqs.~\ref{eqn:line-1}--\ref{eqn:line-4}, this adversely affects the resulting Stokes $Q$ and $U$, where $\sigma^2_0$ enters as a multiplicative term. As the magnetic geometry approaches and crosses the Van Vleck angle, the $\sigma^2_0$ alignment passes through zero and changes sign, producing vanishing linear polarization and additional degeneracies in the retrieved solutions. Thus, without prior knowledge of this sign, we can not discriminate between parallel or perpendicular magnetic fields with respect to the linear polarization measurement. When having access to two-line Stokes $V$ measurements, the alignment correction beyond the standard longitudinal Zeeman contribution (see Eq.~\ref{eqn:line-4}) that can help determine the sign of the alignment, bypassing this degeneracy. The Van Vlack effect for M1 lines like the \ion{Fe}{13} pair has been extensively covered  \citep{querfeld1982,judge2007,Dima+2020,schad+dima2021}. We note that breaking the axial symmetry of the radiation field about the local vertical will alter this interpretation, shifting the zero-polarization condition and introducing variability in its location  \citep{schad+dima2021}.

As mentioned, the degeneracy multiplicity for the selected $IQUV$ or $IQUD$ inversion setups is not the same.  With the exceptions discussed above, the inversion returns pairs of 4-times degenerate solutions, with respect to the observer Cartesian $B_x$ and $B_z$ components. An example of this degeneracy is plotted in Fig.~\ref{fig:degeneracy}, and presented in Tab. \ref{tab:inversion}. The four blue plus red colored degeneracies are given by the unknown sign of $B_x$ (weakly constrained $\sigma^2_0$), and correspond to outputs of the $IQUD$ inversion. Afterward, in the $IQUV$ case, the field strength and sign of the $B_x$ component (whether the field is pointing towards or away from the observer) are inferred through the Stokes $V$ measurement, reducing the degeneracy multiplicity to either the red or blue sets of solutions. In the $IQUD$ setup, the field scaling follows the analytical interpretation given by Eq.~\ref{eqn:phase2}, which is found to be effectively equivalent to using Stokes $V$. The $IQUD$ inversion algorithm does not currently implement any additional constraint on $B \sin \Theta_B$ for multiplicity reduction, as these relaxed assumptions have not yet been sufficiently validated by observations or modeling work. Further analysis is required to prove if additional information can be extracted from coronal seismology analysis to interpret the orientation of $\Theta_B$, before such constraint can be implemented in the CLEDB $IQUD$ setup.

Further, we stress the importance of exploring additional degeneracy reducing schemes: e.g.  physically constrained functionals, analogous to the disambiguation commonly used for photospheric magnetograms \citep[Ambig code;][]{leka2009}; enforcing continuity conditions like $\nabla B=0$; tomography applications \citep{Kramar2006,Kramar+2016}, etc. Any such prospective setups or combinations offer a pathway for breaking at least one degree of degeneracy and obtaining 2-times degenerate solution as analogous to a $IQUV$ inversion, and maybe even towards disambiguated fields. 

%----------------------------------------
\subsection{Discussion on Assumptions and Implications}
\label{ssec:assumptions}

The degeneracies described above are a direct result of solving an inversion problem, that is oftentimes ill-posed or even under-determined for certain geometries. However, these are not the only challenges that manifest when inferring coronal magnetic fields. We highlight a set of important considerations for the observational and model assumptions implicitly embedded in our spectropolarimetric inversion results, and discuss their implications.

\begin{deluxetable}{ccccccccccc}
\tablenum{3}
\tablecaption{\fontsize{10}{12}\selectfont CLEDB $IQUV$ and $IQUD$ inversion solutions for an example forward synthesized $10^{-3}$, perturbed observation of a $1G$ field. Four $\chi^2$ ranked solutions (from $IQUV$ ranking) are presented, along with degeneracy on signs of $B_x$ and $B_z$ colored in blue and red respectively. \label{tab:inversion}}
\tablewidth{0pt}
\tablehead{
\colhead{DB Index} & 
\colhead{$\chi^2$} & 
\colhead{$n_e$} & 
\colhead{$y$} & 
\colhead{$x$} & 
\colhead{$B$} & 
\colhead{$\phi$} & 
\colhead{$\theta$} & 
\colhead{$B_x$} & 
\colhead{$B_y$} & 
\colhead{$B_z$}
}
\startdata
 Obs.   &         & 8.35  & 1.048 & -0.035& 1.000 & 0.786  & 2.338  & 0.513                    & 0.495 & -0.700\\
\tableline
34403   & 0.05746 & 8.352 & 1.050 & -0.040 & 0.998 & 0.768 & 0.803 & \textcolor{blue}{0.516}  & 0.499 & 0.695 \\
34447   & 0.05746 & 8.352 & 1.050 & -0.040 & 0.998 & 0.768 & 2.339 & \textcolor{blue}{0.516}  & 0.499 & -0.695 \\
297743  & 0.05751 & 8.352 & 1.050 & ~0.040  & 0.998 & 2.374 & 0.803 & \textcolor{red}{-0.516} & 0.499 & 0.695 \\
297787  & 0.05751 & 8.352 & 1.050 & ~0.040  & 0.998 & 2.374 & 2.339 & \textcolor{red}{-0.516} & 0.499 & -0.695 \\
\enddata
\tablecomments{\fontsize{10}{12}\selectfont $n_e$ in log cm$^{-3}$, $y$ and $x$ in $R_\odot$, $\phi$ and $\theta$ in radians, and magnetic components in $G$.}
\end{deluxetable}

In the CLEDB inversion framework, the plasma is assumed to be predominantly originating from one dominant location along the LOS. This is called the single-point approximation. Broadly, this assumption has been used for a diverse variety of solar and stellar applications of radiative transfer, plasma diagnostics, and magnetic field diagnostics, \citep[e.g.][]{Rees1979,querfeld1982,querfeld1984,judge1998,phillips2008}, and even in modern model and inversion frameworks \citep{asensioramos2007,ramirez2010,par+judge2022}. We stress that distinct assumption sets and limitations exist for distinct applications; e.g. integrating through optically thick versus optically thin regimes, etc. The single-point approximation represents in general a backbone for prototyping inversion efforts upon which more physically detailed inversion schemes can be designed, where in some cases this assumption can even be ultimately dropped. \citep[e.g.][]{socasnavarro2000,Kramar2006,vannoort2012}. 

In the optically thin off-limb corona, the inversion of Stokes profiles can become challenging when multiple emitting structures contribute along the LOS. In such cases, the CLEDB algorithm ingests Stokes measurements representative of blended spectral features or indistinguishably smeared profiles. \citet{querfeld1982,querfeld1984}, and \citet{judge2007} argued that although most points along a LOS discretization can contribute significantly, the bulk of the emission should cluster within a region $\sim\pm10\%$ around the POS, due to the sharp drop in the IR \ion{Fe}{13} photoexcitation rates with increasing height. Our results support this interpretation, showing that lower heights provide stronger constraints on the inversion. In this context, the CLEDB~V0.7.0 LOS position discretization of $\pm 0.5R_{\odot}$ allows height deviations from the observed $y$ ranging from approximately $\sim\pm 12\%$ at $1R_{\odot}$ to $\sim\pm 3\%$ at $2R_{\odot}$.

But is this reasoning generally valid? The impact of LOS contributions also depends on the spatial resolving capabilities of the instrumentation. Although the optically thin LOS integration itself is unaffected, limited spatial resolution may blend emissions from multiple coronal structures, which in some locations, challenge the validity of the single-point approximation. For instance, assume we attempt inverting active region corona, where two closely spaced and similarly oriented thin loops formed in close temperatures and density regimes. In a low resolution observation, these two loops can still smear into a coherent structure. But if these would be spread far enough along the LOS such that photoexcitation differences becomes significant, or if the magnetic geometries significantly diverge, the resulting integrated Stokes $IQUV$ profiles will fail to invert to a correct result. Although finding that single sources can be fairly elongated along a line of sight ($<0.2R_\odot$), \citet{par+judge2022} strongly argue that multiple contributing sources will detrimentally influence matched solutions. For this reason, it is important in the case of spectrograph observations to at the least check the deviation from a single component Gaussian distribution (or it's derivative in the case of Stokes $V$). Multiple contributions might be mitigated if the structures are not significantly smeared, by separating the emission using multiple Gaussian fiting and then solving for each component. Alternatively, LOS structures can be disentangled through tomographic reconstructions \citep{Kramar+2016,Plowman2021} that could augment polarimetric inversions like CLEDB. Such methods will introduce distinct assumptions beyond the scope of the present work.

In addition, several theoretical considerations must be taken into account when interpreting CLEDB inversion results. As stressed by \citet{delzanna2021,delzanna2025}, accurate and up to date atomic calculations are required in order to adequately infer the main plasma parameters and interpret forward model calculations of coronal emission. A careful balance between detailed atomic calculations and computational costs is needed. Current PyCELp calculations with 85+ CHIANTI \ion{Fe}{13} levels that are used to build a CLEDB database are found to be moderately demanding computation wise. In the future, the reduced models presented by \citet{delzanna2025} could assist in reducing the computational overhead. Also, the CLEDB inversion operates under the assumption of cylindrically symmetric radiative anisotropy. \citet{schad+dima2021} showed that symmetry breaking effects can be significant when observing above active regions, although the effect was found to be minor in their studied case. More effort is required in order to quantitatively understand this effect, but if such symmetry breaking effects become dominant, our ability to map the Van Vleck crossing and $\sigma^2_0$ sign swap becomes a function of the symmetry breaking. The inversion, as currently posed in CLEDB, will not be able to retrieve the needed physics from such an observation. 

Although this work represents a theoretical approach at validating a novel vector field inversion methodology, a number of assumptions, constraints and limitations will manifest when applying CLEDB inversions:

\begin{enumerate}
\renewcommand{\labelenumi}{\roman{enumi}.}

    \item As noted previously, the information content of a single M1 coronal emission line is insufficient to recover the vector magnetic field. Instead, simultaneous observations of two coronal M1 lines are required, preferably from the same ion to avoid uncertainties associated with elemental abundances. This requirement may be relaxed to observations separated by a few minutes, provided that the observed structures remain effectively unchanged during that interval. No current instrument configuration provides this capability. The development of such observational capabilities will be essential for advancing coronal magnetic-field diagnostics. 

    \item This study demonstrates that in both $IQUV$ and $IQUD$ setups, we can at least theoretically invert localized observations up to $2R_{\odot}$, corresponding to current observatory off-limb FOVs. This does not directly translate to an ability to infer physical properties of coronal plasma in the IR \ion{Fe}{13} lines up to those heights, when demanding accurate photometric and polarimetric accuracy. Our perturbation estimations are relative to line strengths, and do not explicitly model instrumental and atmospheric stray-light. Furthermore, the relative radiance of the \ion{Fe}{13} lines varies significantly with height, from a factor of 1.5 at $\sim1.0R_\odot$ to $> 10$ at $\sim2.0R_\odot$ \citep[See Fig.~8 of][]{schad+dima2020}. As a result, reliable application of the line-ratio density diagnostic (i.e. $n_e$) requires sufficiently high-SNR measurements of both lines, the feasibility of which depends on the available spectral, spatial, and temporal resolution. Inaccurate line-ratios can introduce systematic biases into both the $IQUV$ and $IQUD$ inversion schemes. We note a practical $IQUD$ constraint: the same set of observations must be used to compute $n_e$ when deriving B$_{\text{POS}}$ via wave-tracking and in CLEDB inversions, to ensure consistency between the two inversion frameworks.
    
    \item We find that inversion performance generally improves for lower-height and lower-density observations. This behavior likely reflects a combination of factors including: the evolving balance between photoexcitation and collisional excitation; parameter coupling between density, anisotropy, and linear polarization fraction; uncertainties in the underlying atomic data \citep{delzanna2025}; the intrinsically weak Zeeman signals expected in coronal observations; uncertainties associated with wave-based magnetic-field estimates; limitations of the single-point approximation when applied to diffuse coronal structures. Together, these effects appear to contribute to the progressive degradation of inversion performance toward larger projected heights and in more extreme regions of the physical parameter space. Such realizations of the physical parameters should be interpreted with appropriate caution.

    \item $IQUD$ inversions will not observationally be equivalent to $IQUV$ inversions. The current \textit{IQUD} implementation is designed to take as input the outputs (i.e., B$_{\text{POS}}$) of a separate, complex wave-tracking inversion. Such wave-tracking analysis currently requires that magnetic field strength is averaged on timescales of around one hour. This is necessary to analyze and interpret the waves observations with a reasonable degree of statistical confidence \citep{yang2020+}. Thus, under current methodological restrictions, the $IQUD$ setup can be applied only to synoptic use cases, or when longer separated temporal snapshots are sufficient and adequate for a study. No sensitivity to faster evolving or erupting structures exists in the current CLEDB $IQUD$ implementation. 

    \item Relatedly, this study does not propagate uncertainties originating from wave-tracking inversions. These uncertainties affect the field strength derived through the \textit{IQUD} setup. A more rigorous electron density estimate, $n_{e(m)}$, that utilizes the inferred true height of the structure, is produced as a CLEDB inversion output. The more rigorous $n_{e(m)}$ inference, taken together with $B\sin\Theta_B$ constraints, can subsequently be used to refine the interpretation of wave-tracking results, particularly of $c_{km}$.

    % \item As observed from Earth, the LOS and $x$ are almost parallel. For example, an elongation of the angle between the LOS and $x$ at a height $\sim2R_{\odot}$ is $<0.5^\circ$. Any errors introduced with this assumption are minor compared with the other observational and theoretical challenges presented by the problem at hand. This parallax difference does impose a hard limit on the maximum angle discretization in database configurations to be $>1^\circ$.

    % \item Significant attention needs to be paid towards accurately resolving emission line polarization above background scattering polarization levels in the optically thin corona, especially for Stokes Q and U. A user is required to manually validate the inputs, as this assumption is not enforced and can not be treated implicitly by an inversion scheme operating with integrated Stokes measurements.
    
    \item A number of issue flags are implemented in the CLEDB software and described in the software documentation\footnote{\url{https://cledb.readthedocs.io}} to advise a user on potential issues, but we stress that these are not exhaustive.
\end{enumerate}

%----------------------------------------
\section{Conclusion and Future Prospects}\label{sec:concl}

When aggregating the results presented above, we emphasize the importance of the following:

\begin{itemize}
    \item We devise and introduce a novel ``$IQUD$'' inversion method for inferring 3D coronal magnetic fields using the IR \ion{Fe}{13} coronal emission lines, and demonstrate the feasibility of combining $IQU$-only spectropolarimetry with coronal seismology constraints. As stressed above, the interpretation of traveling Alfv\'{e}nic waves in the corona, together with the application of the $IQUD$ method to observations, presents challenges that might reduce the practical advantage over direct Stokes $V$ measurements. Nevertheless, the $IQUD$ inversion method when it can be suitably applied to an observation, can relax photometric requirements and enable even smaller-aperture instruments to probe the three-dimensional coronal magnetic fields that are the root of space weather events.
    
    \item  This work demonstrates that an $IQUD$ inversion algorithm, using Stokes $IQU$ and coronal seismology observations, is method-wise functionally equivalent to the $IQUV$ inversion that was proposed in \citet{par+judge2022}. We find no overarching differences between the results of the $IQUV$ and $IQUD$ inversions. We identify a complementary application range of the two methods, along with some geometry-dependent deviations, although their overall occurrence is low.
    
    \item Consequently, we report a first theoretical validation of core CLEDB functionality, in both $IQUV$ and $IQUD$ inversion setups. The experiment was designed to mitigate discretization errors and demonstrate the applicability of the method. We recognize that constraints of the observational nature exist. This work represents a stepping stone towards building our capabilities for inferring 3D coronal magnetic fields from existing instruments such as DKIST Cryo-NIRSP, DL-NIRSP and MLSO UCoMP and the future COSMO and CORSAIR coronagraphs. 
    
    \item We remind the reader that although the single-point approximation is utilized in solar and general astrophysics research, it should not be considered as universally applicable in the optically thin corona case, and that inversion output accuracy is directly assumes a dominating structure along a LOS integration. An assessment of the applicability of this assumption for each probed location is therefore required. Any statistical characterization of where this approximation holds in the off-limb corona will depend on the prevailing coronal conditions and the solar cycle phase. In addition, we discuss limitations of inversions in context of including detailed physical assumptions, and stress that continued investigation of line integration effects, followed in importance by the non-symmetric anisotropy effects, and to a lesser degree, refinements atomic models remain important to advance our understanding of inversion inferences.
    
    \item Currently, a degeneracy multiplicity difference exists between the 2-times degenerate $IQUV$ and 4-times degenerate $IQUD$ inversion setups. We discussed the typical errors resulting from the over-simplistic assumption that all emission originates from the POS, and emphasize the importance of accounting for localized dominant contributions along the LOS. The $IQUD$ inversion setup allows finding database combinations of $\Theta_B$, $\sigma^2_0$, and $B$ that are compatible with the observed $c_{km}$. Further work on interpreting propagating Alfv\'{e}nic waves is needed to safely implement this assumption and reduce the current degeneracies in the inversion to bring the $IQUV$ and $IQUD$ approaches to an equivalent degeneracy multiplicity. Furthermore, our results indicate that combining Stokes $V$ and Doppler constraints (via coupled $IQUV$ and $IQUD$ inversions) have the potential to directly infer disambiguated magnetic field solution, marking a critical advance.
    
    \item CLEDB is designed and provided as an open-source community oriented software distribution, that is currently in a beta stage of development. Both $IQUV$ and $IQUD$ inversion schemes are implemented and available publicly starting with CLEDB~V0.7.0. 
\end{itemize}

As a last consideration, we note that although optimistic in nature, the results of this experiment should be taken with a grain of salt. We show throughout this work that the coronal IR \ion{Fe}{13} pair suffers from similar kinds of interpretation limitations as other magnetically sensitive lines formed in other layers of the solar atmosphere. More work and additional constraints, applied implicitly in inversion schemes or imposed separately, are required in order to disentangle the magnetism of the optically-thin corona. 

\begin{acknowledgments}
The author gratefully acknowledges the reviewer for their careful assessment, and for the valuable suggestions that improved the clarity and scope of this work, that strengthened the methodological approach of this study. The author appreciates the very productive discussions, proofreads, and idea bouncing with Daniela Lacatus, Thomas Schad, and Philip Judge. The author utilized a Large Language Model \citep{openai2026_gpt53} strictly for assistance with language editing and content clarity. The author was primarily funded for this work by the National Solar Observatory (NSO), a facility of the National Science Foundation (NSF), operated by the Association of Universities for Research in Astronomy (AURA), Inc., under Cooperative Support Agreement number AST-1400405. This research and scientific software development was partly supported by the NASA Heliophysics Division under grant 80NSSC24K0580. The Daniel K. Inouye Solar Telescope (DKIST) is a facility of the NSO. DKIST is located on land of spiritual and cultural significance to Native Hawai’ian people. The use of this important site to further scientific knowledge is done so with appreciation and respect. The author would also like to acknowledge the fruitful interactions with the Mauna Loa Solar Observatory (MLSO) team. MLSO is an observatory facility operated by the High Altitude Observatory, as part of the National Center for Atmospheric Research (NCAR), supported by the NSF.
\end{acknowledgments}

\software{CLEDB~V0.7.0 \citep[Code repository: \href{https://github.com/arparaschiv/solar-coronal-inversion/releases/tag/v0.7.0-beta}{github.com/arparaschiv/solar-coronal-inversion}]{par+judge2022};
PyCELp \citep[Code Repository: \href{https://github.com/tschad/pycelp}{github.com/tschad/pycelp}]{pycelp2021}; 
Astropy \citep{astropy2022};
Sunpy \citep{sunpy_community2020}; 
Numpy \citep{numpy2020}; 
Scipy \citep{SciPy2020}; 
Matplotlib \citep{matplotlib2007}.}

\section*{Data Availability}
While the use of randomized perturbations and a specific CLEDB database construction method might prevent exact replication of the numerical results, the study design is intended to preserve the robustness of the main metrics. To alleviate concern, we additionally make available the randomly perturbed datasets, along with an annotated executable Python Jupyter notebook, sufficient to reproduce the results, figures and tables presented in this work provided the default precompiled CLEDB database is used. The datasets are hosted through the Harvard Dataverse at
\dataset[doi:10.7910/DVN/OANAQQ]{ https://dataverse.harvard.edu/dataset.xhtml?persistentId=doi:10.7910/DVN/OANAQQ}. A local installation of CLEDB V0.7.0+ and a built or downloaded precompiled database are required for running the annotated notebook.
%\appendix
\bibliography{bibliography}{}
\bibliographystyle{aasjournal}
\end{document}